\documentclass[prb,showpacs]{revtex4}
\usepackage{amssymb}
\usepackage{graphics}
\usepackage{graphicx}

\begin{document}

\title{ Electron energy spectrum and density of states for \\
non-symmetric heterostructures in an in-plane magnetic field }
\author{A. Hern{\'{a}}ndez-Cabrera}
\email{ajhernan@ull.es}
\author{P. Aceituno}
\affiliation{Dpto. F{\'\i}sica B\'{a}sica, Universidad de La Laguna, La Laguna,
38206-Tenerife, Spain}
\author{F.T. Vasko}
\email{ftvasko@yahoo.com}
\affiliation{Institute of Semiconductor Physics, NAS Ukraine, Pr. Nauki 41, Kiev, 03028,
Ukraine}
\date{\today}

\begin{abstract}
Modifications of spin-splitting dispersion relations and density of states
for electrons in non-symmetric heterostructures under in-plane magnetic
field are studied within the envelope function formalism. Spin-orbit
interactions, caused by both a slow potential and the heterojunction
potentials (which are described by the boundary conditions) are taken into
account. The interplay between these contributions and the magnetic field
contribution to the spin-splitting term in the Hamiltonian is essential when
energy amount resulting from the Zeeman and spin-orbit coupling are of the
same order. Such modifications of the energy spectra allow us to separate
the spin-orbit splitting contributions due to a slow potential and due to
the heterojunctions. Numerical estimates for selectively-doped
heterojunction and quantum well with narrow-gap region of electron
localization are performed.
\end{abstract}

\maketitle

\section{Introduction}

Spin-orbit splitting of the energy dispersion relations for electrons in
nonsymmetric quantum heterostructures has been theoretically considered
during the past decades (see Ref. \onlinecite{1} for a review). In bulk
materials spin-orbit interaction appears both due to a slow-variable
potential (related to the lattice constant) \cite{2} and due to cubic \cite%
{3} and linear \cite{4} spin-dependent contributions to the effective
Hamiltonian. Turning to the two-dimensional (2D) case, we can reduce the
cubic contribution to a linear one after the replacement of the squared
momentum by the quantized value due to confinement \cite{5}. It is still
more important the fact that we have to take into account an additional
spin-orbit splitting of the energy spectrum due to the interaction with
abrupt heterojunction potentials (see Ref. \onlinecite{6} and discussion in
Refs. \onlinecite{1,7}). Such contribution is of a radically different kind
with respect to the listed above because contributions from both sides of a
slow confinement potential compensate each other \cite{8} and, therefore,
the spin-splitting of 2D states can not be obtained without a short-range
potential contribution. To the best of our knowledge, the relative
contributions from the bulk-induced mechanisms \cite{2,3,4} and from the
heterojunctions have not been clarified experimentally in spite of a set of
existing theoretical calculations \cite{1}. In this paper we have examined
the effect of an in-plane magnetic field on the electron energy spectrum and
density of states and we have found that magneto-induced modifications of
these characteristics are essentially different from the 2D spin-orbit
interaction and Zeeman splitting. In principle, this fact makes possible an
experimental verification of the above discussed contributions of the
spin-orbit interaction.

\begin{figure}[tbp]
\begin{center}
\includegraphics{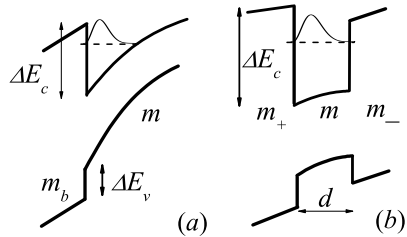}
\end{center}
\par
\addvspace{-1 cm}
\caption{Band diagrams for selectivly-doped heterojunction ($a$) and quantum
well ($b$). Dashed lines show electron ground state energy levels and thin
curves correspond to wave functions; $d$ is the QW's width, $\Delta E_{c,v}$
are the band offsets, and $m$ is the effective mass in the narrow-gap
region, while $m_{b}$ and $m_{\pm }$ are the barrier effective masses.}
\label{Fig. 1}
\end{figure}

The effect of in-plane magnetic field on the energy spectrum in
non-symmetric heterostructures occurs due to the Zeeman term in the electron
Hamiltonian. This fact has been found in Ref. \onlinecite{9}, shortly after
initial considerations of the spin-orbit splitting in non-symmetric
heterostructures (see Refs. \onlinecite{1,10,11}). A number of peculiarities
in transport phenomena for 2D systems under an in-plane magnetic field were
also discussed \cite{11,12,13} and more complicated cases, such as quasi-1D
transport or spin Hall effect (see Refs. in \cite{14} or \cite{15},
respectively) were recently considered. All this papers only take into
account the mix between the Zeeman contribution and the effective 2D
spin-orbit interaction. Here we perform the calculations based on the
three-band Kane model\cite{16} with non-symmetric boundary conditions, which
is valid for narrow-gap heterostructures, and with a slow potential
described self-consistently (see typical band diagrams in Fig. 1). The
analysis of magneto-induced modifications of 2D energy spectra and
corresponding density of states is presented for typical parameters of
InGaAs/InAlAs-selectively-doped heterojunctions, and InGaAs and InSb-based
quantum wells (QWs).

The paper is organized as follows. In the next section we discuss the
conduction $c$-band eigenstate problem and describe the electronic states in
non-symmetric narrow-gap heterostructures using the boundary conditions for
the wave functions at the interfaces in the parabolic approximation. In Sec.
III we solve the 2D-eigenstate problem using the averaged transverse field
approach. Numerical self-consistent calculations of the magneto-induced
(caused by a magnetic field) modifications of the energy spectra and the
density of states are presented in Sec. IV. Conclusions are given in the
last section.

\section{Basic equations}

We start from the formulation of the eigenstate problem for\ the electronic
state of the conduction $c$-band localized in the selectively-doped
heterojunction or in the non-symmetric QW. All the above listed
spin-dependent contributions are taken into account in the analysis. We also
present the density of states which is connected to the photoluminescence
excitation (PLE) intensity.

\subsection{Three-band Kane model}

The electronic states in narrow-gap heterostructures are described by the
three-band Kane matrix Hamiltonian \cite{16} 
\begin{equation}
\hat{\varepsilon}_{z}+(\hat{\mathbf{v}}\cdot \hat{\mbox{\boldmath $\pi$}}%
),~~~~~~\hat{\mbox{\boldmath $\pi$}}=\hat{\mathbf{p}}-\frac{e}{c}\mathbf{A},
\end{equation}%
where the kinetic momentum, $\hat{\mbox{\boldmath $\pi$}}$, contains the
vector potential $\mathbf{A}=(Hz,0,0)$, $\mathbf{H}\Vert OY$ is an in-plane
magnetic field and $\hat{\mathbf{p}}=(\mathbf{p},\hat{p}_{z})$ is written in
the $\mathbf{p},z$-representation through the 2D momentum $\mathbf{p}$. Here
we have also introduced the diagonal energy matrix $\hat{\varepsilon}_{z}$
whose elements determine the positions of the band extrema (the energy
values $\varepsilon _{cz}$, $\varepsilon _{hz}$, and $\varepsilon _{lz}$
correspond to electron, heavy- and light-hole extrema) and the interband
velocity matrix $\hat{\mathbf{v}}$. The $6\times 6$ Hermitian matrix $\hat{%
\mathbf{v}}$ is determined by the following non-zero matrix elements \cite{7}
: 
\begin{eqnarray}
v_{13}^{x} &=&v_{26}^{x}=\mathcal{P}/\sqrt{2},~~~~v_{15}^{x}=-v_{24}^{x}=-%
\mathcal{P}/\sqrt{6},  \nonumber \\
v_{13}^{y} &=&-v_{26}^{y}=-i\mathcal{P}/\sqrt{2},~~~~v_{15}^{y}=v_{24}^{y}=-i%
\mathcal{P}/\sqrt{6},  \nonumber \\
v_{14}^{z} &=&v_{25}^{z}=\sqrt{2/3}\mathcal{P}
\end{eqnarray}%
and $\mathcal{P}$ is the characteristic interband velocity for the Kane
model.

Neglecting the small contributions from other bands we suppose in (1) that
the inverse heavy hole effective mass is equal to zero. For such a case
valence $v$-band components of the wave functions, $\psi _{\mathbf{p}%
z}^{(3-6)}$, are expressed in terms of the $c$-band spinor with components $%
\psi _{\mathbf{p}z}^{(1,2)}$ and the effective Schr\"{o}dinger equation for
this spinor takes the form (see Ref. \onlinecite{17} where the case $H=0$
have been considered): 
\begin{eqnarray}
&&\left\{ \varepsilon _{cz}-E-\frac{2}{3}\mathcal{P}^{2}\left( \frac{\pi
_{+}\pi _{-}}{\varepsilon _{vz}-E}+\hat{p}_{z}[\varepsilon _{vz}-E]^{-1}\hat{%
p}_{z}\right) \right.  \nonumber \\
\left. +\left[ 
\begin{array}{ll}
~~~~0 & \left[ \frac{\pi _{+}}{\varepsilon _{vz}-E},\hat{p}_{z}\right] \\ 
-\left[ \frac{\pi _{-}}{\varepsilon _{vz}-E},\hat{p}_{z}\right] & ~~~~0%
\end{array}%
\right] \right\} \left[ 
\begin{array}{l}
\psi _{\mathbf{p}z}^{(1)} \\ 
\psi _{\mathbf{p}z}^{(2)}%
\end{array}%
\right] &=&0,
\end{eqnarray}%
where $\pi _{\pm }=(\pi _{x}\pm i\pi _{y})$ are the circular components of
the in-plane momenta. For the sake of simplicity we have neglected here the
strain effect due to lattice mismatch \cite{18} and used $\varepsilon
_{vz}\equiv \varepsilon _{hz}=\varepsilon _{lz}$. It is convenient to
introduce in Eq. (3) the$\ z$-dependent effective mass, $m_{z}$, the
characteristic spin velocity, $V_{z}$, and the effective $g$-factor, $g_{z}$%
, according to the relations: 
\begin{eqnarray}
m_{z} &=&-\frac{3}{4\mathcal{P}^{2}}(\varepsilon _{vz}-E),~~~V_{z}=\frac{%
\hbar }{4m_{z}}\frac{(d\varepsilon _{vz}/dz)}{\varepsilon _{vz}-E}, 
\nonumber \\
g_{z} &=&\frac{m_{e}}{2m_{z}}\left[ 1-z\frac{(d\varepsilon _{vz}/dz)}{%
\varepsilon _{vz}-E}\right] .
\end{eqnarray}

In such definitions Eq. (3) may be rewritten as the spinor eigenvalue
problem: 
\begin{eqnarray}
&&\left\{ \varepsilon _{cz}-E+\frac{(p_{x}-eHz/c)^{2}+p_{y}^{2}}{2m_{z}}+%
\hat{p}_{z}\frac{1}{2m_{z}}\hat{p}_{z}\right.  \nonumber \\
\left. -V_{z}\left[ \hat{\mbox{\boldmath $\sigma$}}\times \mathbf{p}\right]
_{z}+\frac{g_{z}}{2}\mu _{\scriptscriptstyle B}H\hat{\sigma}_{y}\right\}
\Psi _{\mathbf{p}z} &=&0,
\end{eqnarray}%
where spinor $\Psi _{\mathbf{p}z}$ is determined by the components $\psi _{%
\mathbf{p}z}^{(1,2)}$, $\mu _{B}\equiv |e|\hbar /(m_{e}c)$ is Bohr magneton,
and $\hat{\mbox{\boldmath $\sigma$}}$ is the Pauli matrix. The energy values
of $c$- and $v$-band extrema for the narrow-gap ($A$) and wide-gap ($B$)
regions take the forms: 
\begin{eqnarray}
\varepsilon _{cz} &=&U_{z},~~~~~~\varepsilon _{vz}=-\varepsilon
_{g}+U_{z},~~~~~~~(A) \\
\varepsilon _{cz} &=&\Delta E_{c}+U_{z},~~~~\varepsilon _{vz}=-\varepsilon
_{g}+\Delta E_{v}+U_{z},~~~~(B)  \nonumber
\end{eqnarray}%
and the corresponding band diagrams for QW and selectively-doped
heterojunction were shown in Fig.1. Here the energy is counted from the
bottom of $c$-band, $\varepsilon _{g}$ is the gap, $\Delta E_{c}$ and $%
\Delta E_{v}$ are the band offsets for $c$- and $v$-bands, correspondingly.
Here the slow potential $U_{z}$ should be determined from a self-consistent
procedure. Thus, after substitution of (6) in Eq. (5) we have formulated the
eigenstate problem for the spinor $\Psi _{\mathbf{p}z}$.

Since $\Delta E_{c}\geq \varepsilon _{g}$ in narrow-gap heterostructures,
then the weak underbarrier penetration of wave function takes place for the
electronic states with $E\ll \varepsilon _{g}$ (parabolic band
approximation). Because of this, one can neglect the longitudinal motion for
the underbarrier region, $z<0$ for the single heterojunction case shown in
Fig.1$(a)$, where the solutions take the form: 
\begin{equation}
\Psi _{\mathbf{p}z}\simeq \Psi _{\mathbf{p}z=0}e^{\kappa z},~~~~~z<0.
\end{equation}%
Here $\hbar \kappa =\sqrt{2m_{b}\Delta E_{c}}$ and $\kappa ^{-1}$determines
the scale of the underbarrier penetration of wave functions. We have also
used in Eq. (7) the continuity conditions for eigenfunctions $\Psi _{\mathbf{%
p}z}$ at heterojunctions. Since $V_{z}$ in Eq. (4) is proportional to $%
(d\varepsilon _{vz}/dz)$, then the integration (5) over heterojunction
produces additional contributions to the boundary condition for flows. Such
contributions are proportional to the band offset at the heterojunction $%
\Delta E_{v}$. Eliminating the underbarrier contributions from such equation
by the use of the explicit expression (7) we obtain the third kind boundary
conditions \cite{6,7}: 
\begin{equation}
(\hat{p}_{z}\Psi _{\mathbf{p}z})|_{z=0}-iP_{o}\Psi _{\mathbf{p}z=0}-i\chi %
\left[ \hat{\mbox{\boldmath $\sigma$}}\times \mathbf{p}\right] _{z}\Psi _{%
\mathbf{p}z=0}=0.
\end{equation}%
Here the momenta $P_{o}=\hbar \kappa m/m_{b}$ characterizes the underbarrier
penetration and the parameter $\chi $ determines the spin-orbit coupling due
to the abrupt potential of the heterojunction 
\begin{equation}
\chi =\frac{2m_{b}}{\hbar }\int_{-\delta }^{\delta }dzV_{z}\simeq \frac{%
\Delta E_{v}}{2\varepsilon _{g}}
\end{equation}%
and the right-side part is written for the approximation $m_{b}\simeq m$

\subsection{Eigenstate problem}

In the parabolic approximation, we describe the $c$-band electronic states
using the above introduced spinor $\Psi _{\mathbf{p}z}$. Below we consider
not very strong magnetic fields ($\hbar \omega _{c}\ll v_{\scriptscriptstyle %
F}\hbar /d$, where $v_{\scriptscriptstyle F}$ is the Fermi velocity and $%
\omega _{c}$ is the cyclotron frequency) when $(p_{x}-eHz/c)^{2}$ in Eq. (5)
is replaced by $p_{x}^{2}$ and the isotropic kinetic energy is given by $%
\varepsilon _{p}\equiv (p_{x}^{2}+p_{y}^{2})/(2m)$ and includes the
effective mass $m$. Considering the low-doped structure case (if $%
\varepsilon _{g}\gg \varepsilon _{\scriptscriptstyle F}$) we rewrite the Schr%
\"{o}dinger equation (5) for the narrow-gap region in the form: 
\begin{eqnarray}
\left( \varepsilon _{p}+\frac{\hat{p}_{z}^{2}}{2m}+U_{z}+\widehat{W}%
_{z}\right) \Psi _{\mathbf{p}z} &=&E\Psi _{\mathbf{p}z},  \nonumber \\
\widehat{W}_{z} &=&-V_{z}\left[ \hat{\mbox{\boldmath $\sigma$}}\times 
\mathbf{p}\right] _{z}+\frac{g_{z}}{2}\mu _{\scriptscriptstyle B}H\hat{\sigma%
}_{y}
\end{eqnarray}%
where $U_{z}$ is the self-consistent potential,\ as given in Eq. (14) below,
and the effective mass is $z$-independent. The spin velocity $V_{z}$ and the 
$z$-dependent contribution to the $g$-factor are proportional to the
transverse electric field $dU_{z}/dz$: 
\begin{equation}
V_{z}\simeq -\frac{\hbar }{4m\varepsilon _{g}}\frac{dU_{z}}{dz}%
,~~~~~~g_{z}\simeq \frac{m_{e}}{2m}\left( 1+\frac{z}{\varepsilon _{g}}\frac{%
dU_{z}}{dz}\right) .
\end{equation}%
For the selectively-doped heterojunction case we consider Eq. (10) for the $%
z>0$ region with the boundary conditions (8) and with $\Psi _{\mathbf{p}%
z\rightarrow \infty }=0$. For the case of a QW of width $d$, one can
eliminate the underbarrier contributions in analogy to Eqs. (7-9) and, in
addition to Eq. (10) for the region $|z|<d/2$, the following third kind
boundary conditions should be used: 
\begin{equation}
(\hat{p}_{z}\Psi _{\mathbf{p}z})|_{z=\pm \frac{d}{2}}\mp iP_{\pm }\Psi _{%
\mathbf{p},\pm \frac{d}{2}}+i\chi _{\pm }\left[ \hat{%
\mbox{\boldmath
$\sigma$}}\times \mathbf{p}\right] _{z}\Psi _{\mathbf{p},\pm \frac{d}{2}}=0.
\end{equation}%
Here momenta $P_{\pm }=\sqrt{2m_{\pm }\Delta E_{\pm }}(m/m_{\pm })$
determine the scale of the underbarrier penetration of wave functions; the
different values of $\Delta E_{\pm }$ and $m_{\pm }$ take into account the
differences of band offsets and effective masses, as it is shown in Fig. 1$%
\left( b\right) $. The parameters $\chi _{\pm }\simeq \Delta
E_{v}/2\varepsilon _{g}$ determine the spin-orbit coupling due to the abrupt
potential of the heterojunction according to Eq. (9). When $m_{+}=m_{-}$ ,
which corresponds to the symmetric case, we will take $P_{\pm }=P_{o}$ and $%
\chi _{\pm }=\chi $, as we can see from (9).

Thus, both interface potentials, which determine the spin-dependent
contributions to (12), and the intrawell field, which determines the spin
velocity $V_{z}$ in (11), are responsible for the spin-splitting of energy
spectra. Since the Zeeman spin-splitting term only appears in the Eq. (10),
the mentioned magneto-induced modifications of electron states due to these
two contributions are different.

The self-consistent numerical procedure for the eigenstate problem (10)
involves the potential $U_{z}$, which is obtained from the Poisson equation
in the following form: 
\begin{equation}
U_{z}=\frac{4\pi e^{2}}{\epsilon }\int_{-\infty }^{z}dz^{\prime }\mathbf{(}%
z-z^{\prime }\mathbf{)}[n_{\scriptscriptstyle D}(z^{\prime
})-n_{e}(z^{\prime })]
\end{equation}%
Here $n_{\scriptscriptstyle D}(z)$ is the 3D concentration of donors and $%
\epsilon $ is the dielectric permittivity that we have supposed as uniform
across the heterostructure. The electron density distribution is introduced
through the electron dispersion relations $E_{\sigma \mathbf{p}}$ according
to: 
\begin{equation}
n_{e}(z)=\sum_{\sigma }\int \frac{d\mathbf{p}}{(2\pi \hbar )^{2}}\Psi _{%
\mathbf{p}z}^{{\scriptscriptstyle(\sigma )+}}\cdot \Psi _{\mathbf{p}z}^{%
\scriptscriptstyle(\sigma )}\theta (\varepsilon _{\scriptscriptstyle %
F}-E_{\sigma \mathbf{p}}),
\end{equation}%
where $\Psi _{\mathbf{p}z}^{{\scriptscriptstyle(\sigma )+}}$ denotes the
Hermitian conjugate of $\Psi _{\mathbf{p}z}^{{\scriptscriptstyle(\sigma )}}$
and \ $\sigma =\pm 1$ refers to the two possible spin orientations. The
Heaviside function, $\theta (x)$, appears here for the zero-temperature
case. The Fermi energy, $\varepsilon _{\scriptscriptstyle F}$, is expressed
through the total electron density, $n_{\scriptscriptstyle2D}$, defined as $%
n_{\scriptscriptstyle2D}=\int dzn_{e}(z).$ Thus, $\varepsilon _{%
\scriptscriptstyle F}$ depends on $H$ for the fixed concentration case.

The density of states is given by the standard formula: 
\begin{equation}
\rho _{\varepsilon }=\sum_{\sigma }\int \frac{d\mathbf{p}}{(2\pi \hbar )^{2}}%
\delta (\varepsilon -E_{\sigma \mathbf{p}}).
\end{equation}%
In order to analyze $\rho _{\varepsilon }$ one needs to solve the
above-formulated eigenstate problem and to perform the integrations in Eq.
(15). The density of states is connected to PLE intensity for the case of
near-edge transitions, $I_{\scriptscriptstyle PLE}$. Since the interband
matrix element $\mathbf{v}_{cv}$ do not depend on the in-plane quantum
numbers, one obtains \cite{7,19}: 
\begin{equation}
I_{\scriptscriptstyle PLE}\sim \sum_{\lambda _{c}\lambda _{v}}|\mathbf{%
e\cdot v}_{cv}|^{2}\delta (\varepsilon _{\lambda _{c}}-\varepsilon _{\lambda
_{v}}-\hbar \omega )\sim \rho _{\hbar \Delta \omega },
\end{equation}%
where $\mathbf{e}$ is the polarization vector, $\Delta \omega =\omega -%
\overline{\varepsilon }_{g}/\hbar $ and $\overline{\varepsilon }_{g}$ is the
gap energy, which is renormalized due to the confinement effect.

\section{Analytical consideration}

Before the numerical consideration we perform a simplified calculation of
the eigenstate problem using the uniform transverse field approximation $%
U_{z}\simeq eF_{\scriptscriptstyle\bot }z$ which is valid for non-doped QWs
under an external modulating field $F_{\scriptscriptstyle\bot }$; for
heavy-doped structures $F_{\scriptscriptstyle\bot }$ implies an averaged
self-consistent transverse field.

\subsection{Average field approach}

A simplification of the eigenstate problem (10) appears due the to $z$%
-independent spin-orbit perturbation when $\widehat{W}_{z\text{ }}$in Eq.
(10) becomes 
\begin{equation}
\widehat{W}=\bar{\mathrm{v}}\left[ \hat{\mbox{\boldmath $\sigma$}}\times 
\mathbf{p}\right] _{z}+w_{\scriptscriptstyle H}\hat{\sigma}_{y}
\end{equation}%
with the averaged across structure characteristic spin velocity $\overline{%
\mathrm{v}}\simeq |e|F_{\scriptscriptstyle\bot }\hbar /(4m\varepsilon _{g})$
and the Zeeman splitting $w_{\scriptscriptstyle H}=(\bar{g}/2)\mu _{%
\scriptscriptstyle B}H$, where the $z$-dependent correction to the $g$%
-factor is neglected. The fundamental solutions $\Psi _{\mathbf{p}z}$ for
the Eq. (10), with the spin-dependent contribution (17), can be factorized
as products of functions $\Psi _{\mathbf{p}}^{\scriptscriptstyle(\sigma )}$
and $\varphi _{z}^{(k\sigma )}$. Here the spinors $\Psi _{\mathbf{p}}^{%
\scriptscriptstyle(\sigma )}$ for $\sigma =\pm 1$ are determined from the
eigenstate problem: 
\begin{equation}
\left( \varepsilon _{p}+\widehat{W}\right) \Psi _{\mathbf{p}}^{%
\scriptscriptstyle(\sigma )}=\varepsilon _{\sigma \mathbf{p}}\Psi _{\mathbf{p%
}}^{\scriptscriptstyle(\sigma )},
\end{equation}%
whose solutions are given \cite{10} by: 
\begin{eqnarray}
\Psi _{\mathbf{p}}^{\scriptscriptstyle(+1)} &=&\frac{1}{\sqrt{2}}\left[ 
\begin{array}{l}
~~~~~~~~~~~1 \\ 
(\overline{\mathrm{v}}p_{+}+w_{\scriptscriptstyle H})/i\mathrm{w}_{\mathbf{p}%
}%
\end{array}%
\right] ~~,  \nonumber \\
\Psi _{\mathbf{p}}^{\scriptscriptstyle(-1)} &=&\frac{1}{\sqrt{2}}\left[ 
\begin{array}{l}
(\overline{\mathrm{v}}p_{-}+w_{\scriptscriptstyle H})/i\mathrm{w}_{\mathbf{p}%
} \\ 
~~~~~~~~~~~1%
\end{array}%
\right] ~~, \\
\varepsilon _{\sigma \mathbf{p}} &=&\varepsilon _{p}+\sigma \mathrm{w}_{%
\mathbf{p}},~~~~~\mathrm{w}_{\mathbf{p}}=\sqrt{(\bar{\mathrm{v}}p_{x}+w_{%
\scriptscriptstyle H})^{2}+(\overline{\mathrm{v}}p_{y})^{2}},  \nonumber
\end{eqnarray}%
where $p_{\pm }=p_{x}\pm ip_{y}$ and the energy values $\varepsilon _{\pm 
\mathbf{p}}$ describe the mix between internal spin-orbit interaction and
Zeeman spin splitting.

The two $z$-dependent fundamental solutions $\varphi _{z}^{\scriptscriptstyle%
(k\sigma )}$ (labeled below by $k=a,b$) are determined from the equation 
\begin{equation}
\left( \frac{\hat{p}_{z}^{2}}{2m}+eF_{\scriptscriptstyle\bot }z\right)
\varphi _{z}^{\scriptscriptstyle(k\sigma )}=\left( E-\varepsilon _{\sigma 
\mathbf{p}}\right) \varphi _{z}^{\scriptscriptstyle(k\sigma )},
\end{equation}%
written for the narrow-gap region.

The general spinor solution $\Psi _{\mathbf{p}z}$ is expressed through these
fundamental solutions according to 
\begin{eqnarray}
\Psi _{\mathbf{p}z} &=&\Psi _{\mathbf{p}}^{\scriptscriptstyle(+1)}\left(
A_{+}\varphi _{z}^{\scriptscriptstyle(a,+1)}+B_{+}\varphi _{z}^{%
\scriptscriptstyle(b,+1)}\right)  \nonumber \\
&&+\Psi _{\mathbf{p}}^{\scriptscriptstyle(-1)}\left( A_{-}\varphi _{z}^{%
\scriptscriptstyle(a,-1)}+B_{-}\varphi _{z}^{\scriptscriptstyle(b,-1)}\right)
\end{eqnarray}%
where the coefficients $A_{\pm },B_{\pm }$ are determined from the boundary
conditions (12) or from (8) and the requirements $\Psi _{\mathbf{p}%
z\rightarrow \infty }=0$.

After the substitution of the solution (21) for $\varphi _{z}^{%
\scriptscriptstyle(b\sigma )}=0$, which corresponds to the boundary
condition at $z\rightarrow \infty $, into (8) and the multiplication of this
system by $\Psi _{\mathbf{p}}^{\scriptscriptstyle(\sigma )}$ on the left
side we rewrite the boundary condition at $z=0$ as follows: 
\begin{eqnarray}
\mathcal{P}_{+}A_{+}+\chi \Psi _{\mathbf{p}}^{\scriptscriptstyle(+1)+
}{}\cdot \left[ 
\begin{array}{ll}
0 & -p_{+} \\ 
p_{-} & 0%
\end{array}%
\right] \Psi _{\mathbf{p}z=0} &=&0,  \nonumber \\
\mathcal{P}_{-}A_{-}+\chi \Psi _{\mathbf{p}}^{\scriptscriptstyle(-1)+
}{}\cdot \left[ 
\begin{array}{ll}
0 & -p_{+} \\ 
p_{-} & 0%
\end{array}%
\right] \Psi _{\mathbf{p}z=0} &=&0,
\end{eqnarray}%
with $\mathcal{P}_{\scriptscriptstyle\pm }\equiv \left( \hat{p}%
_{z}+iP_{o}\right) \varphi _{z}^{\scriptscriptstyle(a\pm )}\left\vert
_{z=0}\right. $. In order to calculate the proportional to $\chi $
contributions we use here the solutions of the spin-dependent eigenstate
problem (19). Thus, Eq. (22) can be transformed with the use of the
relations 
\begin{eqnarray}
\Psi _{\mathbf{p}}^{(\sigma )}{}^{\scriptscriptstyle +}\cdot \left[ 
\begin{array}{ll}
0 & -p_{+} \\ 
p_{-} & 0%
\end{array}%
\right] \Psi _{\mathbf{p}}^{(\sigma )} &=&\sigma \frac{\overline{\mathrm{v}}%
p^{2}+w_{\scriptscriptstyle H}p_{x}}{iw_{\mathbf{p}}}  \nonumber \\
\Psi _{\mathbf{p}}^{(\sigma )}{}^{\scriptscriptstyle + }\cdot \left[ 
\begin{array}{ll}
0 & -p_{+} \\ 
p_{-} & 0%
\end{array}%
\right] \Psi _{\mathbf{p}}^{(-\sigma )} &=&-i\frac{w_{\scriptscriptstyle %
H}p_{y}}{\mathrm{v}p_{\sigma }+\sigma w_{\scriptscriptstyle H}}
\end{eqnarray}%
to a simple linear system for $A_{\pm }$. The dispersion relation, $%
E_{\sigma \mathbf{p}}$, is determined from the zero determinant requirement.

A similar transformation of the boundary conditions (12) for QWs, after
substitution on (21), permit us to rewrite the boundary condition at $z=\pm
d/2$ in the form: 
\begin{eqnarray}
&&\left( \hat{p}_{z}\mp iP_{\pm }\right) \left( A_{+}\varphi _{z}^{%
\scriptscriptstyle(a,+1)}+B_{+}\varphi _{z}^{\scriptscriptstyle%
(b,+1)}\right) \left\vert _{z=\pm \frac{d}{2}}\right.  \nonumber \\
+\chi \Psi _{\mathbf{p}}^{\scriptscriptstyle(+1)+ }{}\cdot \left[ 
\begin{array}{ll}
0 & -p_{+} \\ 
p_{-} & 0%
\end{array}%
\right] \Psi _{\mathbf{p},\pm \frac{d}{2}} &=&0,  \nonumber \\
&&\left( \hat{p}_{z}\mp iP_{\pm }\right) \left( A_{-}\varphi _{z}^{%
\scriptscriptstyle(a,-1)}+B_{-}\varphi _{z}^{\scriptscriptstyle%
(b,-1)}\right) \left\vert _{z=\pm \frac{d}{2}}\right.  \nonumber \\
+\chi \Psi _{\mathbf{p}}^{\scriptscriptstyle(-1)+ }{}\cdot \left[ 
\begin{array}{ll}
0 & -p_{+} \\ 
p_{-} & 0%
\end{array}%
\right] \Psi _{\mathbf{p},\pm \frac{d}{2}} &=&0.
\end{eqnarray}%
Thus, we have obtained a linear system for $A_{\pm }$ and $B_{\pm }$ and,
therefore, $E_{\sigma \mathbf{p}}$ is obtained from the solvability
condition for this system.

\subsection{2D model}

First, let us consider the case of a heterostructure without spin-orbit
contributions from heterojunctions, $\chi =0$, when the dispersion relation
is given by Eq. (19). For the zero magnetic field case, $w_{%
\scriptscriptstyle H}=0$, the energy $\mathrm{w}_{\mathbf{p}}$ is replaced
by $\overline{\mathrm{v}}p$ and $\varepsilon _{\pm \mathbf{p}}$ in Eq. (19)
is transformed to the isotropic dispersion relation $\varepsilon _{\sigma
p}=\varepsilon _{p}+\sigma |\overline{\mathrm{v}}|p$. For this case the
density of states was considered in Ref. \onlinecite{7}. If $w_{%
\scriptscriptstyle H}\neq 0$, the dispersion relation (19) becomes
anisotropic as it is shown in Fig. 2 $(a)$. We have used in this figure the
dimensionless magnetic field $h=w_{\scriptscriptstyle H}/m\overline{\mathrm{v%
}}^{2}$. In order to represent a general case, valid for any structure, we
have also used dimensionless energy axis.

\begin{figure}[tbp]
\begin{center}
\includegraphics{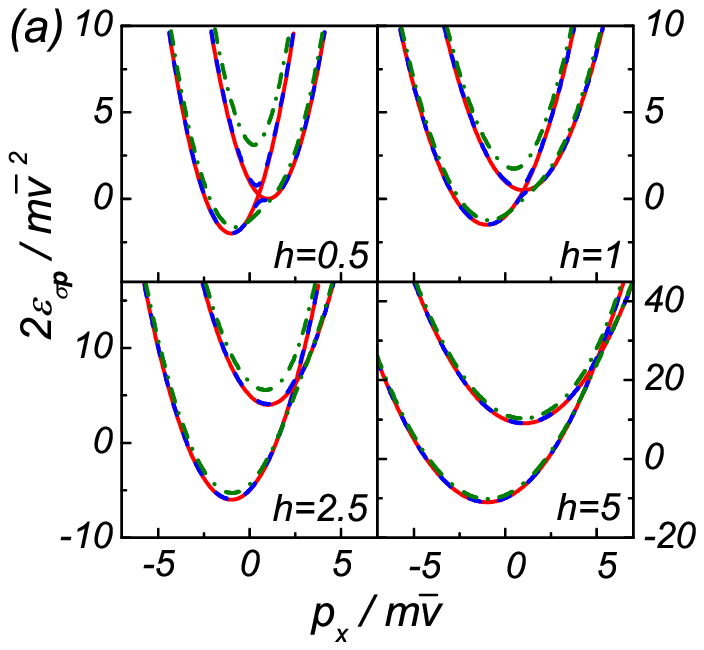} \includegraphics{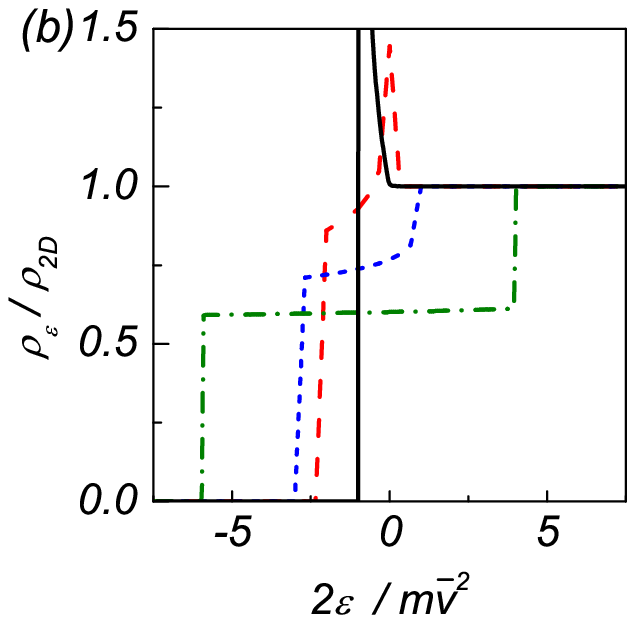}
\end{center}
\par
\addvspace{-1 cm}
\caption{(Color on line) (a) Dispersion laws for different $h$ values.
Solid, dashed and dot-dashed curves corresponds to the dimensionless momenta 
$p_{y}/m\overline{\mathrm{v}}=$0, 0.25, and 1, respectively. (b) Density of
states vs dimensionless energy for magnetic fields $h=$0 (solid), 0.5
(dashed), 1 (dotted), 2.5 (dash-dotted), and 5 (dash-dot-dotted). }
\end{figure}

For the case of the strong magnetic field, when $\mathrm{w}_{\mathbf{p}}$ is
replaced by $w_{\scriptscriptstyle H}$, the Zeeman splitting effect appears
to be dominant in the dispersion relation: $\varepsilon _{\pm p}\simeq
\varepsilon _{p}\pm w_{\scriptscriptstyle H}$. The spin-orbit splitting is
mainly manifested as a shift of the dispersion paraboloids towards higher
(lower) $p_{x}$ values for $\sigma =+1(-1)$, respectively. On the other
hand, Zeeman splitting is shown as a displacement to higher (lower) energy
values depending on $\sigma $. The anisotropy of the dispersion relation $%
\varepsilon _{\sigma \mathbf{p}}$ appears to be essential for the region
around the cross-point , $\varepsilon _{c}\equiv w_{\scriptscriptstyle %
H}^{2}/2m\overline{\mathrm{v}}^{2}$.

After the shift $p_{x}\rightarrow p_{x}-w_{\scriptscriptstyle H}/m\overline{%
\mathrm{v}}$ and the integration over $p$ with the use of the $\ \delta $%
-function, the density of states is transformed from Eq. (15) into the
integral with respect to the cosine of the in-plane angle, $\zeta =\cos
\theta $: 
\begin{eqnarray}
\frac{\rho _{\varepsilon }}{\rho _{\scriptscriptstyle2D}} &=&\frac{1}{2\pi }%
\sum_{\sigma }\int_{-1}^{1}\frac{d\zeta }{\sqrt{1-\zeta ^{2}}} \\
&&\times \frac{\theta (\nu _{\sigma \zeta }^{2}-1+2\varepsilon /hw_{%
\scriptscriptstyle H})}{\sqrt{\nu _{\sigma \zeta }^{2}-1+2\varepsilon /hw_{%
\scriptscriptstyle H}}}\sum_{\pm }\mathcal{E}_{\pm }\theta (\mathcal{E}_{\pm
}).  \nonumber
\end{eqnarray}%
Here $\mathcal{E}_{\pm }=\nu _{\sigma \zeta }\pm \sqrt{\nu _{\sigma \zeta
}^{2}-1+2\varepsilon /hw_{\scriptscriptstyle H}}$ and $\nu _{\sigma \zeta
}=\zeta -\sigma /h$. A straightforward integration for the above-cross-point
region $\varepsilon >\varepsilon _{c}$ gives as the exact relation: $\rho
_{\varepsilon }=\rho _{\scriptscriptstyle2D}$, while for the
below-cross-point region $\varepsilon <\varepsilon _{c}$ we plot $\rho
_{\varepsilon }/\rho _{\scriptscriptstyle2D}$ versus dimensionless energy $%
2\varepsilon /m\overline{\mathrm{v}}^{2}$ for different fields $h$. Figure 2$%
(b)$ shows this density of states for different dimensionless magnetic
fields. Two points deserve special attention. First, the typical $%
\varepsilon ^{-1/2}$ singularity around $\varepsilon =0$, which occurs for $%
h=0$, tends to disappears with increasing field. Second, abrupt steps
corresponding to the filling of each level are shifted with the field
intensity indicating a delay in the filling caused by the Zeeman splitting.
It should be noted that the lowest $\left( \sigma =-1\right) $ level seems
to be overfilled for low magnetic fields, i.e., $\rho _{\varepsilon }/\rho _{%
\scriptscriptstyle2D}>1/2$. Electrons equally distributes between levels for
magnetic fields higher than $h=10$.

\section{Numerical results}

Here we perform numerical calculations based on the simplified consideration
outlined in Sec. II$B$, as well as the self-consistent solution of the
eigenstate problem of Sec. II$A$. The correspondent level-splitting at the
Fermi energy level in a selectively-doped heterojunction and the densities
of states in QWs are described.

\subsection{Selectively-doped heterojunction}

The eigenfunction of Eq. (20) for $z>0$ is written through the \ Airy $Ai$%
-function , $Ai[z/l_{\bot }-(E-\varepsilon _{\sigma \mathbf{p}})/\varepsilon
_{\bot }]$ with $l_{\bot }=\sqrt[3]{\hbar ^{2}/2m|e|F_{\bot }}$ and $%
\varepsilon _{\bot }=(\hbar /l_{\bot })^{2}/2m$, if the zero boundary
condition at $z\rightarrow \infty $ is applied. After substitution of this
function into the system (22), one obtains the dispersion equation in the
form: 
\begin{equation}
\det \left[ 
\begin{array}{ll}
\mathcal{P}_{\scriptscriptstyle+}-i\chi \frac{\overline{\mathrm{v}}p^{2}+w_{%
\scriptscriptstyle H}p_{x}}{w_{\mathbf{p}}}\varphi _{0}^{\scriptscriptstyle%
(a,+1)} & ~~~~-i\chi \frac{w_{\scriptscriptstyle H}p_{y}}{\overline{\mathrm{v%
}}p_{+}+w_{\scriptscriptstyle H}}\varphi _{0}^{\scriptscriptstyle(a,-1)} \\ 
~~~~-i\chi \frac{w_{\scriptscriptstyle H}p_{y}}{\overline{\mathrm{v}}%
p_{-}+w_{\scriptscriptstyle H}}\varphi _{0}^{\scriptscriptstyle(a,+1)} & 
\mathcal{P}_{\scriptscriptstyle-}+i\chi \frac{\overline{\mathrm{v}}p^{2}+w_{%
\scriptscriptstyle H}p_{x}}{w_{\mathbf{p}}}\varphi _{0}^{\scriptscriptstyle%
(a,-1)}%
\end{array}%
\right] =0.
\end{equation}%
After calculating this determinant and introducing the dimensionless
function $K(x)\equiv -Ai^{\prime }(x)/Ai(x)$ one transforms Eq. (26) into 
\begin{eqnarray}
&&\left[ \frac{P_{o}l_{\bot }}{\hbar }+K\left( \frac{\varepsilon _{{%
\scriptscriptstyle+1}p}-E}{\varepsilon _{\bot }}\right) -v_{s}\frac{%
\overline{\mathrm{v}}p^{2}+w_{\scriptscriptstyle H}p_{x}}{\varepsilon _{\bot
}w_{\mathbf{p}}}\right]  \nonumber \\
&&\times \left[ \frac{P_{o}l_{\bot }}{\hbar }+K\left( \frac{\varepsilon _{{%
\scriptscriptstyle-1}p}-E}{\varepsilon _{\bot }}\right) +v_{s}\frac{%
\overline{\mathrm{v}}p^{2}+w_{\scriptscriptstyle H}p_{x}}{\varepsilon _{\bot
}w_{\mathbf{p}}}\right]  \nonumber \\
-\left( \frac{v_{s}w_{\scriptscriptstyle H}p_{y}}{\varepsilon _{\bot }w_{%
\mathbf{p}}}\right) ^{2} &=&0,~~~
\end{eqnarray}%
where $v_{s}=\chi \varepsilon _{\bot }l_{\bot }/\hbar $ is the spin velocity
due to the interface contribution. The spin-dependent dispersion relation, $%
E_{\sigma \mathbf{p}}$, is given by the roots of this equation.

Let us consider the 2D approach, when $E$ is around the ground energy level, 
$\varepsilon _{o}$, determined by the equation: $P_{o}l_{\bot }/\hbar
+K(-\varepsilon _{o}/\varepsilon _{\bot })=0$. Expanding $K[(\varepsilon _{{%
\sigma }p}-E)/\varepsilon _{\bot }]$ over $E-\varepsilon _{o}-\varepsilon
_{p}$ one can transform Eq. (27) into a quadratic equation and the
dispersion relation takes the form $E_{\sigma \mathbf{p}}\simeq \varepsilon
_{o}+\varepsilon _{p}+\sigma \Delta E_{\mathbf{p}}$, where the splitting $%
\Delta E_{\mathbf{p}}$ of the energy spectrum is given by 
\begin{equation}
\Delta E_{\mathbf{p}}=\sqrt{\left( w_{\mathbf{p}}-\overline{v}_{s}\frac{%
\overline{\mathrm{v}}p^{2}+w_{\scriptscriptstyle H}p_{x}}{w_{\mathbf{p}}}%
\right) ^{2}+\left( \frac{\overline{v}_{s}w_{\scriptscriptstyle H}p_{y}}{w_{%
\mathbf{p}}}\right) ^{2}}.
\end{equation}%
with the heterojunction-induced spin velocity, $\overline{v}%
_{s}=v_{s}/K(-\varepsilon _{o}/\varepsilon _{\bot })$, and the internal spin
velocity, $\overline{\mathrm{v}}$, introduced in Eq. (17). In Fig. 3 we plot
these dispersion relations for the same parameters used in Fig. 2$(a)$. As
in Fig. 2$(a)$, we have used dimensionless energy and momenta to represent a
general case. Comparing Fig. 2$\left( a\right) $ with Fig. 3 we can see the
effect of the interface contribution as an enhancement of the splitting
between parabolas $\sigma =\pm 1$.

\begin{figure}[tbp]
\begin{center}
\includegraphics{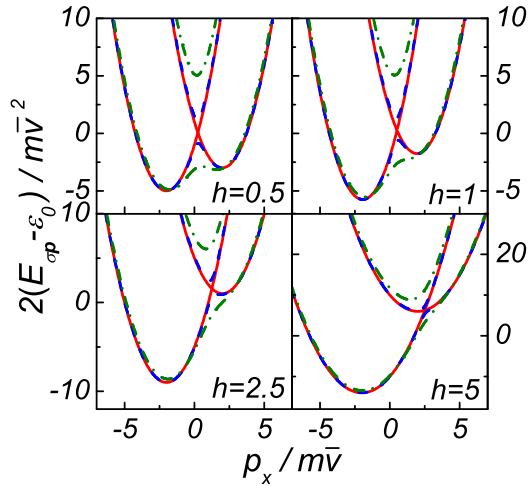}
\end{center}
\par
\addvspace{-1 cm}
\caption{(Color online) Energy dispersion relations calculated with equation
(28) for $\overline{v}_{s}=\overline{\mathrm{v}}$ for different magnetic
fields $h.$ Solid, dashed and dot-dashed curves corresponds to the
dimensionless momenta $p_{y}/m\overline{\mathrm{v}}=$0, 0.25, and 1,
respectively.}
\end{figure}

For high electron concentration the approach of Eq. (28) is no longer valid.
In order to obtain $E$ for this case we need to solve Eq. (10) together with
Eq. (13) by means of a self-consistent approach. To do that we have used the
Transfer Matrix Method\cite{20}. Figure 4 shows dispersion relations
obtained in this way for In$_{0.8}$Al$_{0.2}$As/In$_{0.52}$Ga$_{0.48}$As
with the electron density $n_{\scriptscriptstyle2D}=10^{12}$ cm$^{-2}$,
which corresponds to a Fermi energy $\varepsilon _{F}=209$ meV. Using the
standard parameters \cite{21} one obtains $\chi =0.303$ and $\overline{%
\mathrm{v}}=1.13\times 10^{6}$ cm/s, so that dimensionless field $h=1$
corresponds to $H=0.23$ T. Thus, we can compare panel for $h=1$ in Fig.3
with panel for $H=0.23$ T in Fig. 4. We can see that numerically calculated
results show a similar behavior than approximation (28), although, for the
material under consideration, splitting is qualitatively less and shift
along momentum is larger in Fig. 4.

\begin{figure}[tbp]
\begin{center}
\includegraphics{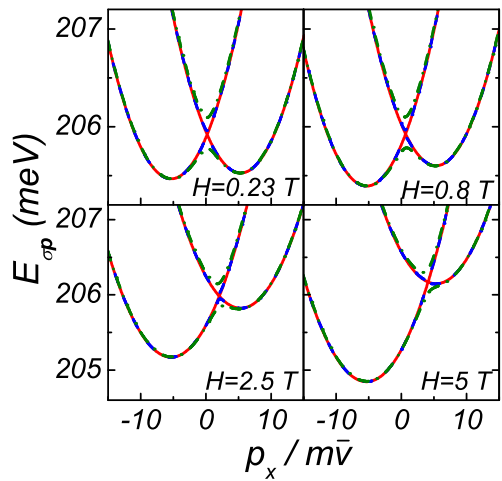}
\end{center}
\par
\addvspace{-1 cm}
\caption{(Color online) Dispersion relations calculated numerically for
different magnetic field $H$ and $n_{2D}=10^{12}$ cm$^{-2}.$ Solid line: $%
p_{y}=0$. Dashed line: $p_{y}=0.25$ $m\overline{\mathrm{v}}.$ Dot-dashed
line: $p_{y}=m\overline{\mathrm{v}}$. }
\end{figure}

The spin splitting of the dispersion relations at Fermi energy is directly
connected to the Shubnikov-de Haas (SdH) oscillations. Since $\Delta E_{%
\mathbf{p}}$ is nearly isotropic over $\mathbf{p}$-plane, we have also
calculated $\Delta E=\Delta E_{\mathbf{p}}|_{p=p_{\scriptscriptstyle F}}$
for different magnetic fields, see Fig. 5. Following Refs. \onlinecite{12,13}
the spin splitting is related to the modulation of the SdH oscillation
amplitude according to $\cos (\pi m\Delta E/\hbar |e|H)$. For the structure
under consideration, experimental value\cite{13} is $\Delta E=11.4$ meV for $%
H=0.8$ T and $n_{\scriptscriptstyle2D}=10^{12}$ cm$^{-2}$, which corresponds
to $\Delta E=9.86$ meV, calculated numerically for the same magnetic field
and density. Figure 5 shows a very slight dependence of $\Delta E$ on the
magnetic field for $\chi =0.303$ being stronger this dependence for the case 
$\chi =0$. Thus, interface contributions should be detectable through the
SdH oscillations amplitude.

\begin{figure}[tbp]
\begin{center}
\includegraphics{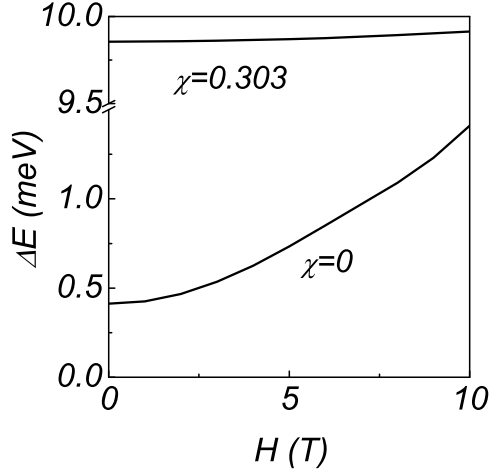}
\end{center}
\par
\addvspace{-1 cm}
\caption{The level splitting energy, $\Delta E$ vs magnetic field for $%
\protect\chi =0.303$ and $\protect\chi =0$. Note that the upper $\Delta E$
is almost $H$-independent. }
\end{figure}

\subsection{QW under field $F_{\scriptscriptstyle \bot}$}

Next, we consider the non-doped symmetric QW (with $P_{\pm }=P_{o}$ and $%
\chi _{\pm }=\chi $) subjected to a transversal field $F_{\bot }$, when the
eigenstate functions of Eq. (20), $\varphi _{z}^{(a,b)}$, are written
through the $Ai$- and $Bi$-functions. The system (24) gives the dispersion
relations equation as the determinant of a $4\times 4$ matrix, which differs
from Eq. (26) due to the replacement of $\mathcal{P}_{\pm }$ and $\varphi
_{o}^{\pm }$ by the $2\times 2$ matrices: 
\begin{equation}
\hat{\mathcal{P}}_{\sigma }=\left[ 
\begin{array}{ll}
\mathcal{P}_{\scriptscriptstyle+}^{\scriptscriptstyle(a\sigma )} & \mathcal{P%
}_{\scriptscriptstyle+}^{\scriptscriptstyle(b\sigma )} \\ 
\mathcal{P}_{\scriptscriptstyle-}^{\scriptscriptstyle(a\sigma )} & \mathcal{P%
}_{\scriptscriptstyle-}^{\scriptscriptstyle(b\sigma )}%
\end{array}%
\right] ,~~~\hat{\varphi}_{\sigma }=\left[ 
\begin{array}{ll}
{\varphi }_{d/2}^{\scriptscriptstyle(a\sigma )} & {\varphi }_{d/2}^{%
\scriptscriptstyle(b\sigma )} \\ 
{\varphi }_{-d/2}^{\scriptscriptstyle(a\sigma )} & {\varphi }_{-d/2}^{%
\scriptscriptstyle(b\sigma )}%
\end{array}%
\right]
\end{equation}%
where $\mathcal{P}_{\scriptscriptstyle\pm }^{\scriptscriptstyle(k\sigma
)}\equiv \left( \hat{p}_{z}\varphi _{z}^{(k,{\scriptscriptstyle\pm 1}%
)}\right) _{z=\pm d/2}\mp iP_{o}\varphi _{\pm d/2}^{(k,{\scriptscriptstyle%
\pm 1})}.$ 
\begin{figure}[tbp]
\begin{center}
\includegraphics{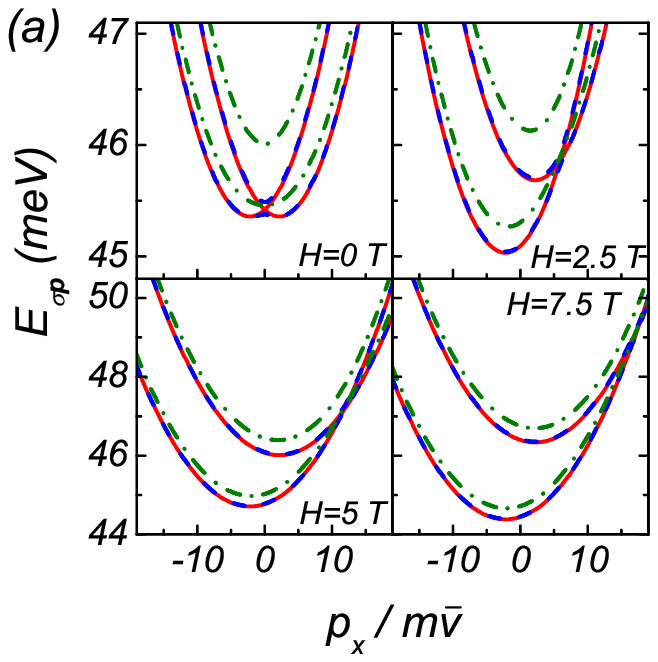}\includegraphics{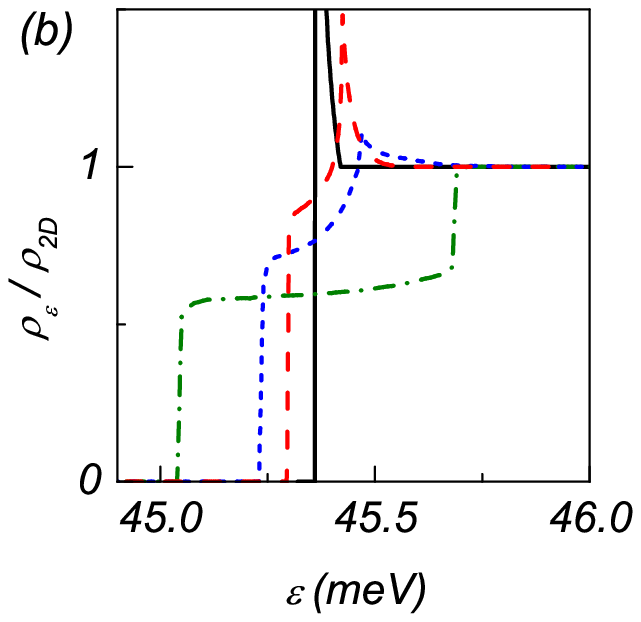}
\end{center}
\par
\addvspace{-1 cm}
\caption{(Color online) (a) Dispersion laws for InGaAs/InAlAs QW under
different magnetic fields. $p_{y}=0$ (solid line) and $p_{y}=m\overline{%
\mathrm{v}}$ (dashed line). (b) Density of states for $H=0$ $T$\ (solid
line), $0.5$ $T$ (dashed line), $1$ $T$ (dot line), and $2.5$ $T$
(dot-dashed line).}
\end{figure}
Once again, after the expansion near the ground energy, $\varepsilon _{o}$,
one can transform the determinant in a similar way to transformations of
Eqs. (26, 27). Numerical solutions of these equations for InGaAs- and
InSb-based QWs are represented in Figs. 6$\left( a\right) $ and 7$\left(
a\right) $. We have considered $100$ \AA\ wide InGaAs- and InSb-based QWs
using the parameters of Refs. \onlinecite{21,22}. We have also applied a
field of $120$ $kV/cm$, which corresponds to spin velocity $\overline{%
\mathrm{v}}=1.04\times 10^{6}$ cm/s, for the InGaAs QW, and a field of $100$ 
$kV/cm$, spin velocity$\ \overline{\mathrm{v}}=8.2\times 10^{6}$ cm/s, for
the InSb.

Comparing Figs. 6$\left( a\right) $ and 7$\left( a\right) $ with Fig. 4 we
can see that, for similar spin velocities and magnetic fields, the splitting
between levels is bigger whereas the shift in the $p_{x}$ direction is
smaller for the QW case with respect to the selectively-doped structure.
Thus, the effect of interface contribution seems to be more pronounced for
the QW case. This effect is enhanced as the width of the wells diminishes.
Due to the greater spin velocities of InGaAs and InSb-based QWs, caused by
the different effective masses and energy gaps, it is necessary the use of
lesser magnetic fields to get a similar effect. Both for QWs and selectively
doped structures, interface contributions are opposite to the intrinsic
spin-orbit coupling effect.

\begin{figure}[tbp]
\begin{center}
\includegraphics{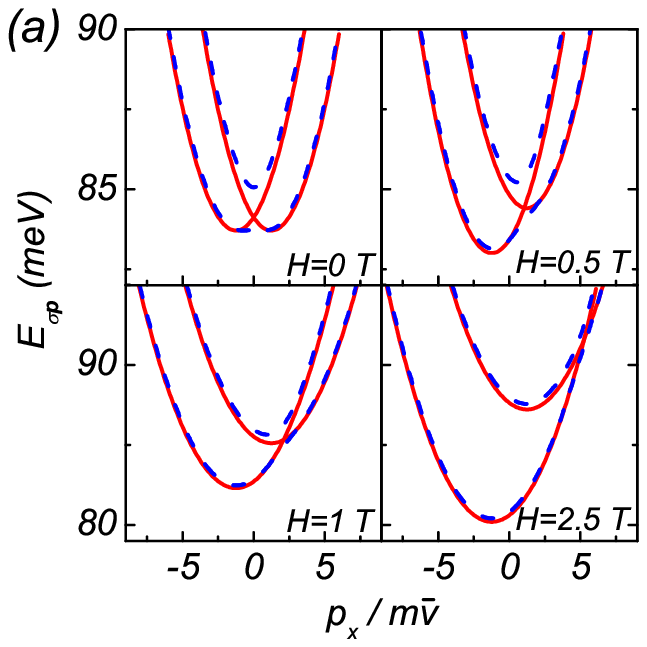}\includegraphics{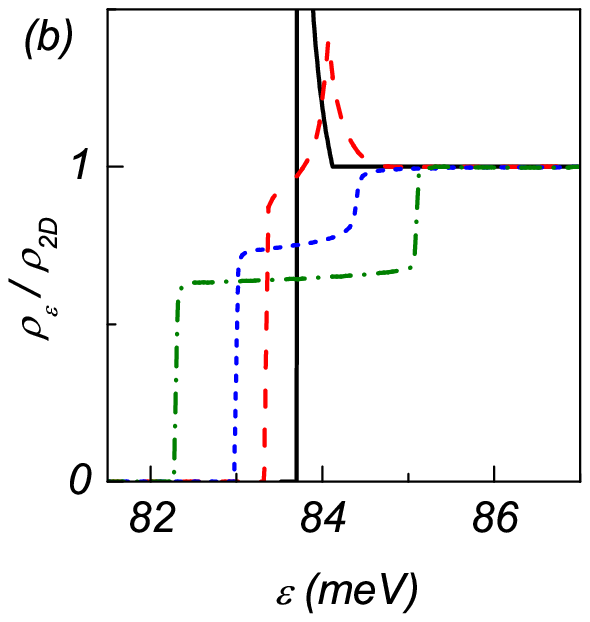}
\end{center}
\par
\addvspace{-1 cm}
\caption{(Color online) (a) Dispersion relations for InSb/InAlSb QW under
different magnetic fields. $p_{y}=0$ (solid line) and $p_{y}=m\overline{%
\mathrm{v}}$ (dashed line). (b) Density of states for $H=0$ $T$\ (solid
line), $0.25$ $T$ (dashed line), $0.5$ $T$ (dotted line), and $1$ $T$
(dot-dashed line).}
\end{figure}

Since the density of states is proportional to the PLE intensity\cite{7, 19}%
, as shown in Eq. (16), it is interesting to study the shape of $\rho
_{\varepsilon }$, shown in Figs. 6$\left( b\right) $ and 7$\left( b\right) $%
. The effect of the interfaces is manifested in $\rho _{\varepsilon }$ as a
delay in the quenching of the $\varepsilon ^{-1/2}$-singularity
corresponding to the zero-field case. Although the singularity no longer
exists for $H\neq 0$, a peak still remains at the energy value of the bands
anticrossing. For the case which includes interface contributions, this peak
gets wider and $\rho _{\varepsilon }$ tends slower to $\rho _{%
\scriptscriptstyle2D}$ for high energy limit. Thus, the PLE technique is of
great interest to analyze the interface contributions\cite{23}. It should be
notice that $\sigma =\pm 1$ levels becomes equally filled for magnetic
fields beyond 5 T.

\section{Conclusion}

In this paper we have examined the electron states in narrow-gap
non-symmetric heterostructures under an in-plane magnetic field. The
eigenstate problem was formulated in the framework of the three-band Kane
model with non-symmetric boundary conditions. We have found that the
mechanisms of mixing between the Zeeman term in the Hamiltonian and the two
kinds of spin-orbit coupling contributions (from a slow field and from
heterojunctions) are essentially different. Numerical estimates for typical
parameters of InGaAs/InAlAs \ and InSb/InAlSb structures demonstrate the
essential magneto-induced modifications of the energy spectra under magnetic
field strength of the order of Tesla.

Let us discuss some possibilities for the experimental verification of the
energy spectra modifications obtained here. The magnetotransport
measurements of Shubnikov-de Haas oscillations under nearly in-plane
magnetic fields\cite{12,13} (when a quasi-classic quantization of the
dispersion relations for the transverse component of the magnetic field is
possible) provide a direct information about in-plane magneto-induced
modifications of electron energy spectra. The results of Sec. IV$A$ are in
agreement with Ref. \onlinecite{13} but more measurements (for different
in-plane fields and concentrations) are necessary in order to separate the
intrinsic and junction-induced contributions. Another way of looking for
these peculiarities of energy spectra is the mid-infrared PLE spectroscopy%
\cite{23} when interband transitions are modified under in-plane magnetic
field. PLE intensity provides direct information of the energy spectra
because it is directly connected to the density of states, as mentioned
above. A comparison between SdH oscillations and PLE measurements with a
precision about 1 meV (it should be possible for a high-quality structure)
provides more data on mechanisms of spin-orbit interaction in narrow-gap
structures.

Thus, we have shown the essential modifications of the electron energy
spectra in non-symmetric narrow-gap QWs under in-plane magnetic fields. We
suggest that measurements of the magneto-induced contributions to optical
and transport properties would be an useful method for an experimental
verification of the heterojunction-induced contributions to the spin-orbit
interaction in the narrow-gap heterostructures under investigation.

\bigskip

\begin{acknowledgments}
This work has been supported in part by Ministerio de Educaci\'{o}n y
Ciencia (Spain) and FEDER under the project FIS2005-01672.
\end{acknowledgments}


\begin{thebibliography}{99}
\bibitem{18} The final Schr\"{o}dinger Eq. (10) and corresponding boundary
conditions (8, 12) do not change under taking into account the
stress-induced splitting of $v$-band extrema and such a revision of the
parameters changes the numerical estimation only slightly but the main
results are the same.
\end{thebibliography}
\end{document}